\documentstyle{ioplppt}                % use this for journal style
\begin{document}
\jl{1}
\title{SU(2) coherent state path integrals
 based on arbitrary fiducial vectors 
and geometric phases\\}[SU(2) coherent state path integrals]
\setcounter{footnote}{0}
\author
{Masao Matsumoto
\footnote{E-mail: matumoto@i.h.kyoto-u.ac.jp}
\footnote{WWW: http://www.i.h.kyoto-u.ac.jp/\~{}matumoto/}}
\address
{1-12-32 Kuzuha Asahi, Hirakata, Osaka 573-1111, Japan}
\date{\today}
\begin{abstract}
We develop the formulation of the spin(SU(2)) coherent state 
path integrals based on arbitrary fiducial vectors. 
The resultant action in the path integral expression 
extensively depends on the vector; 
It differs from the conventional one 
in that it has a generalized form having some additional terms. 
We also study, as physical applications, the geometric phases 
associated with the coherent state path integrals 
to find that new effects of the terms may appear in experiments. 
We see that the formalism gives a clear insight 
into geometric phases.
\end{abstract}
\pacs{03.65.Ca, 03.65.Bz}
\submitted
%\maketitle
%
\section{Introduction} 
It has been more than a quarter of a century since the coherent state 
(CS) for Heisenberg-Weyl group (canonical CS) was extended to wider classes 
\cite{Rad,Pera,Arec,BG}. 
During the period, they, together with the original one, 
have had a great 
influence on almost every branch of modern physics 
\cite{KlaSk,Perb,Perc,IKG,FKS}.
\par
Since basic properties of CS are that they are continuous functions 
labeled by some parameters and that they compose overcomplete sets
\cite{KlaSk}, 
they provide a natural way to perform path integrations. 
Such `coherent state path integrals'(CSPI) 
 have highly enriched 
the methods of path integrals with their physical applications 
\cite{KlaSk,IKG}. 
(In what follows each of the words `CS' and `CSPI' is used as a plural 
as well as a singular.)
\par 
In the present paper, we try to let the method of CS and CSPI 
take another step further; 
Following the recent theoretical development of CSPI 
using the canonical CS \cite{NECSPI}, 
we aim to liberate spin CS from the conventional choice 
of fiducial vectors $\left\vert {\Psi_0} \right\rangle$
\footnote
{
In \cite{NECSPI,spinCSPI} we adopted the term `starting vector' 
which can be found in, e.g. p~14 of \cite{Perb}. 
The term seems well fit for the situation. 
However, we use `fiducial vector' in the present paper 
since it appears to be more employed in literature. 
See e.g. \cite{KlaSk}.
}
 and perform the path integration via the CS 
based on arbitrary fiducial vectors. 
The reason for doing such an extension is twofold. 
First, describing geometric phases, 
which is one of the current topics in fundamental physics 
for more than a decade \cite{SW}, 
in terms of CSPI requires the extension. 
Let us put it more concretely: 
Elsewhere we have investigated the geometric phases 
of a spin-$s$ particle under a magnetic field 
in the formalism of SU(2)CSPI with the conventional 
fiducial vector, i.e. $\left\vert {s, -s} \right\rangle$ \cite{KMa}. 
In consequence the results give the geometric phase 
of a monopole-type that merely corresponds 
to the adiabatic phase for the lowest eigenstate. 
However, it has been known that in the adiabatic phase 
the strength of a fictitious monopole 
is proportional to the quantum number 
$m$ ($ m = -s, - s + 1, \cdots, s$) of the adiabatic state \cite{Berry}. 
We cannot treat the case by the conventional SU(2)CSPI. 
Therefore the usual SU(2)CSPI is clearly unsatisfactory; 
And we had better let CS and CSPI prepare room also 
for the general cases which are reduced to any $m$th eigenstate 
in the adiabatic limit. 
Thus physics actually needs some extension. 
Second, since spin CS tends to the canonical CS in the high spin limit, 
we are led to seek the spin CSPI that is contracted to 
the canonical CSPI with arbitrary fiducial vectors 
described in \cite{NECSPI}. 
Hence we take a general fiducial vector in this paper. 
%
%Second, the extension may also find its way more into physics; 
%Making full use of the arbitrariness of fiducial vectors, 
%it will prove particularly useful to study the collective motions 
%that cannot be treated by CS with the conventional fiducial vectors. 
%
\par
The main results are as follows: 
The form of the generic Lagrangian for the SU(2)CSPI is 
(\ref{eqn:Lagpol}). 
It has the monopole-like term whose strength is proportional 
to the expectation value of the 
quantum number $m$ in the state of $\left\vert {\Psi_0} \right\rangle$; 
And besides (\ref{eqn:Lagpol}) contains additional terms 
that reflect the effect of interweaving coefficients 
of $\left\vert {\Psi_0} \right\rangle$ with their next ones. 
The geometric phases associated with the spinCS in 2-form yields 
(\ref{eqn:2form-pol}); 
It gives a stronger result (\ref{eqn:Gam2}) which, 
in an adiabatic case, reduces to that of \cite{Berry}. 
Thus we can demonstrarte that 
{
\em the SU(2)CSPI is 
mature and complete enough to incorporate the formula 
{\rm (\ref{eqn:Gam2})} 
as its special case 
and moreover it also covers the wider cases in 
$\S$ \ref{sec:simp-s1} - $\S$ \ref{sec:manif-s1}.
}
\par
The plan of the paper is as follows.  
We describe the spin CS based on arbitrary fiducial vectors 
as well as its various properties 
($\S$ \ref{sec:CS}) and 
employ them to perform path integration 
($\S$ \ref{sec:PI}). 
Next we study, as applications, 
 the effect of the CS on the problem of geometric phases 
 to point out new possibilities of their experimental detections 
($\S$ \ref{sec:CSGP}). 
Section \ref{sec:summary} is devoted to the summary and prospects. 
We add \ref{sec:Rots} that serves as the mathematical tools 
of proving the relations in $\S$ \ref{sec:CS} - $\S$ \ref{sec:PI}. 
\section{Coherent state with general fiducial vectors}
\label{sec:CS} 
In this section we investigate the explicit form of SU(2)CS 
based on arbitrary fiducial vectors. 
And their properties are studied to such extent as we need later. 
The results in $\S$ \ref{sec:CS} - $\S$ \ref{sec:PI} include 
those for the conventional SU(2)CS 
\cite{Rad,Arec,Perb} and their CSPI 
\cite{Klau,KUS}; 
The latter follow from the former 
when we put $c_{s}=1$ and $c_m=0\ (m \ne s)$, 
or $c_{-s}=1$ and $c_m=0\ (m \ne -s)$ in later expressions. 
% A. Construction of the Coherent State
%
\subsection{Construction of the coherent state}
The SU(2) or spin CS are constructed from the Lie algebra satisfying 
$
{\bf {\hat S}} \times {\bf {\hat S}} = \i \, {\bf {\hat S}}
$ 
where ${\bf {\hat S}} \equiv (\hat S_1,\hat S_2,\hat S_3)$ 
is a matrix vector composed of the spin operators. 
The operators ${\bf {\hat S}}$ are also the infinitesimal operators 
of the irreducible representstion $R^{(s)}(g)$ of $SO(3)$. 
Since $SU(2) \simeq SO(3)$ locally, 
we can also use $SO(3)$ to construct the SU(2) CS.
The SU(2)CS is defined by operating a rotation operator 
with Euler angles 
$\bf\Omega \equiv (\phi, \theta, \psi)$
\footnote
{
Hereafter we adapt the abbreviation 
$\bf\Omega \equiv (\phi, \theta, \psi)$ 
from Radcliffe \cite{Rad} to describe a set of Euler angles 
which specifies the spin CS.
}, 
which is the operator of $R^{(s)}(g)$, 
on a fixed vector(`fiducial vector') in the Hilbert space of $R^{(s)}(g)$ 
\cite{Rad,Pera,Arec}.
\begin{equation}
\fl
\left\vert {\bf\Omega} \right\rangle 
\equiv \left\vert {\phi, \theta, \psi} \right\rangle 
= {\hat R}({\bf\Omega}) \left\vert {\Psi_0} \right\rangle
= \exp (- \i \phi {\hat S_3}) \exp (- \i \theta {\hat S_2})
 \exp (- \i \psi{\hat S_3}) \left\vert {\Psi_0} \right\rangle. 
\label{eqn:CS1}
\end{equation}
In the conventional choice, $\left\vert {\Psi_0} \right\rangle$ 
is taken as 
$\left\vert {s, -s} \right\rangle$ 
or $\left\vert {s, s} \right\rangle$ 
\cite{Rad,Arec}. 
CS with such fiducial vectors are closest to the classical states 
and have 
various useful properties. We appreciate them truly. 
According to the general theory of the CS, however, 
we have much wider possibility in choosing a fiducial vector; 
And in fact it permits any normalized fixed vector 
in the Hilbert space \cite{Pera,KlaSk,Perb}. 
Thus we can take $\left\vert {\Psi_0} \right\rangle$ as
\begin{equation}
\left\vert {\Psi_0} \right\rangle 
= \sum_{m=-s}^{s} c_m \left\vert {m} \right\rangle 
\qquad {\rm with} \qquad
\sum_{m=-s}^{s} \vert {c_m} \vert^2 = 1.
\label{eqn:statvec}
\end{equation}
Hereafter $\left\vert {m} \right\rangle$ stands for 
$\left\vert {s, m} \right\rangle$. 
The fiducial vector will bring us all the information in later sections 
as far as the general theory. 
Looking at the problem in the light of physical applications, 
we need to take an approapriate $\left\vert {\Psi_0} \right\rangle$, 
i.e. 
$\{ c_m \}$, for each system being considered. 
Notice that the reduction of the number of Euler angles 
is not always possible for an arbitrary 
$\left\vert {\Psi_0} \right\rangle$. 
Hence we use a full set of three Euler angles and proceeed 
with it in what follows, which seems suitable for later discussions. 
\footnote
{
The results in \cite{spinCSPI} should be changed into those 
of the present paper 
since the reasoning employed in Appndix B in \cite{spinCSPI} 
is not correct; 
For any $s$, $\left\vert {\Psi_0} \right\rangle$ 
is not necessarily reached 
from $\left\vert {m} \right\rangle$ via $R^{(s)}({\bf\Omega}).$ 
}  
\par
Having written $\left\vert {\Psi_0} \right\rangle$ 
in the form of (\ref{eqn:statvec}), 
SU(2)CS is represented by a linear combination 
of a set of the vectors $\{ \left\vert {m} \right\rangle \}$ as
\numparts
\begin{equation}
\left\vert {\bf\Omega} \right\rangle 
= \sum_{m=-s}^{s} c_m \left\vert {{\bf\Omega}; m} \right\rangle
\label{eqn:CS3a}
\end{equation}
with
\begin{equation} 
\fl \left\vert {{\bf\Omega}; m} \right\rangle 
\equiv \sum_{m'=-s}^{s}\,  
R_{m'{}m}^{(s)} 
({\bf\Omega})
\ \left\vert {m'} \right\rangle 
= \sum_{m'=-s}^{s}\, 
\exp[-\i(m \phi + m' \psi)] \ r_{m'{}m}^{(s)} 
(\theta)\ \left\vert {m'} \right\rangle. 
\nonumber 
\label{eqn:CS3b}
\end{equation}
\endnumparts
See \ref{sec:Rots} (i) for the definitions of 
$R_{m'{}m}^{(s)}$ and $r_{m'{}m}^{(s)}$.
The form of (\ref{eqn:CS3a})-(\ref{eqn:CS3b}) is 
valuable for later arguments.
\par
The state $\left\vert {\bf\Omega} \right\rangle$ may be named 
`extended spin CS', 
yet we will call it just `the CS' in this paper 
since there have been some arguments about the choice of 
such a fiducial vector 
\cite{KlaSk,Perb} 
and the CSPI 
\cite{ToKlau}.
\footnote 
{
It is reviewed in \cite{Ska}. 
The authors of \cite{ToKlau} constructed `universal propagator' 
for various Lie group cases, 
being independent of the representations, 
which yields a different action from ours.
}
We take a simple strategy for the SU(2)CSPI evolving from 
arbitrary fiducial vectors and 
the related geometric phases. 
And we will give their explicit form, 
which it seems has not been given so far. 
%
% B. Resolution of Unity
%
\subsection{Resolution of unity}
\label{sec:resol}
The most important property that the CS enjoy is the 
`overcompleteness relation' or `resolution of unity' 
which plays a central role in performing the path integration. 
It is expressed as 
\begin{equation}
{2s + 1 \over 8\pi^2}
\int \d{\bf\Omega} 
\vert {\bf\Omega} \left\rangle 
\right\langle {\bf\Omega} \vert 
= {\bf 1} 
\qquad
{\rm with}
\qquad 
\d{\bf\Omega} \equiv \sin\theta \d\theta \d\phi \d\psi.
\label{eqn:polresolution}
\end{equation}
For simplicity, we have neglected the 
difference between an integer $s$ and a half-integer $s$, 
which is not essential. 
Concerning the proof, 
there is an abstract way making full use of Schur's lemma 
\cite{Pera,KlaSk,Perb}. 
However, we propose proving it by a slightly concrete method 
which is a natural extension of that used for the original spin CS 
\cite{Rad,Arec}. 
For it indicates clearly what is to be changed when we use a general 
$\left\vert {\Psi_0} \right\rangle$.

\par
{\it Proof.} 
We see from (\ref{eqn:CS3a})-(\ref{eqn:CS3b})
$
\left\langle {\bf\Omega} \right\vert
= \sum_{{\tilde m} = -s}^{s}
\sum_{m''=-s}^{s}\, 
c_{\tilde m}^{*} 
\left( R_{m''{}{\tilde m}}^{(s)}({\bf\Omega}) \right)^{*} 
 \left\langle {m''} \right\vert.
$
Then, with the aid of (\ref{eqn:ortho}) we have 
\begin{eqnarray}
\fl 
\int \left\vert {\bf\Omega} \right\rangle \d{\bf\Omega} 
\left\langle {\bf\Omega} \right\vert 
 = \sum_{m=-s}^{s} \sum_{{\tilde m}=-s}^{s} 
c_m c_{\tilde m}^{*}  \Bigl\{ \sum_{m'=-s}^{s} 
\sum_{m''=-s}^{s}
\Bigl[  \int_0^{\pi} \d\theta \ \sin\theta 
\nonumber \\ 
\times \int_0^{2\pi} \d\phi \int_0^{2\pi} \d\psi 
\left( R_{m''{} {\tilde m}}^{(s)} ({\bf\Omega}) \right)^{*} \, 
R_{m'{}m}^{(s)}({\bf\Omega}) 
\Bigr] \left\vert {m'} \right\rangle 
\left\langle {m''} \right\vert \Bigr\} 
\nonumber \\
\lo= \sum_{m=-s}^{s} \sum_{{\tilde m}=-s}^{s}
c_m c_{\tilde m}^{*} 
\Bigl( 
\sum_{m'=-s}^{s} \sum_{m''=-s}^{s} 
{8\pi^2 \over 2s + 1}\ \delta_{m'', m'} \delta_{{\tilde m}, m} 
\left\vert {m'} \right\rangle
 \left\langle {m''} \right\vert 
\Bigr) 
\label{eqn:proofPolresol}\\
\lo= {8\pi^2 \over 2s + 1} \, 
\sum_{m=-s}^{s} 
c_m c_m^{*} 
\Bigl( \sum_{m'=-s}^{s} 
\vert {m'} \left\rangle 
\right\langle {m'} \vert 
\Bigr)
= {8\pi^2 \over 2s + 1} 
( \sum_{m=-s}^{s} 
\vert {c_m} \vert^2 )\ {\bf 1}
= {8\pi^2 \over 2s + 1} \ {\bf 1}
\nonumber
\end{eqnarray}
which is exactly we wanted. 
\hfill$\Box$
%
% C. Overlap of two Coherent States
%
\subsection{Overlap of two coherent states}
\label{sec:overlap}
The overlap of two CS 
$
\left\vert {{\bf\Omega}_l} \right\rangle 
\equiv 
\left\vert {\theta_l, \phi_l, \psi_l} \right\rangle 
= \sum_{m_l = -s}^{s} c_{m_l} 
\left\vert {\theta_l, \phi_l; m_l} \right\rangle \  (l = 1, 2)
$
is one of those important quantities which we employ 
for various calculations in the CS. It can be derived, 
with the help of (\ref{eqn:CS1}), (\ref{eqn:s1/2}) 
and (\ref{eqn:rotinverse}), as 
\begin{eqnarray}
\fl
\left\langle {{\bf\Omega}_2} 
\vert 
{{\bf\Omega}_1} \right\rangle 
 = \sum_{m_1=-s}^{s} \sum_{m_2=-s}^{s}
c_{m_1} c_{m_2}^{*} 
\left\langle m_2 \right\vert 
{\hat R}(-\psi_2, -\theta_2, -\phi_2)  
{\hat R}(\phi_1, \theta_1, \psi_1)
\left\vert {m_1} \right\rangle \nonumber \\
\lo= \sum_{m_1=-s}^{s} \sum_{m_2=-s}^{s} 
c_{m_1} c_{m_2}^{*} 
R_{m_2{}m_1}^{(s)}(\varphi, \vartheta, \chi) 
     \\
\lo= \sum_{m_1=-s}^{s} \sum_{m_2=-s}^{s} 
c_{m_1} c_{m_2}^{*} 
\exp [- \i (\varphi m_2+\chi m_1)]\ 
r_{m_2{}m_1}^{(s)}(\vartheta).
\nonumber
\label{eqn:overlap1}
\end{eqnarray}
Here 
$r_{m_2{}m_1}^{(s)}(\vartheta)$ is given in \ref{sec:Rots} (i)
 and $(\varphi, \vartheta, \chi)$ is determined by (\ref{eqn:tworots}). 
It is easy to see that any state 
$\left\vert {\bf\Omega} \right\rangle$ is normalized to unity, 
as conforms to our construction of the CS. 
\subsection{Typical matrix elements}
Typical matrix elements that we encounter in later sections 
are: 
\begin{equation}
\fl 
\left\{
\begin{array}{l}
\left\langle {\bf\Omega} \right\vert 
{\hat S}_{3} 
\left\vert {\bf\Omega} \right\rangle 
= A_0(\{ c_m \}) \cos\theta    
-  A_1(\psi; \{ c_m \}) \sin\theta 
\nonumber \\
\left\langle {\bf\Omega} \right\vert 
{\hat S}_{+} 
\left\vert {\bf\Omega} \right\rangle 
= A_0(\{ c_m \}) \sin\theta  \exp(\i \phi)  
+ A_2({\bf\Omega}; \{ c_m \})  
= \left\langle {\bf\Omega} \right\vert 
{\hat S}_{-} 
\left\vert {\bf\Omega} \right\rangle^{*}
\end{array}
\right.
\label{eqn:mat-pol} 
\end{equation}
where
$
\hat S_{\pm} = \hat S_1 \pm \i \hat S_2 
$
and 
\begin{equation}
\fl 
\left\{
\begin{array}{l}
A_0(\{ c_m \}) 
=\sum_{m=-s}^{s} m \vert {c_m} \vert^2  
\\
A_1(\psi; \{ c_m \})  
= \frac12 \sum_{m=-s+1}^{s} f(s, m) 
[ c_m^{*} c_{m-1} \exp(\i \psi) + c_m c_{m-1}^{*} 
\exp(- \i \psi) ]
\\
A_2({\bf\Omega}; \{ c_m \})
= \frac12 \sum_{m=-s+1}^{s} f(s, m) 
\{ (1 + \cos\theta) \exp[\i (\phi + \psi)] c_m^{*} c_{m-1} 
\\ 
\qquad \qquad - (1 - \cos\theta) \exp[\i (\phi - \psi)] c_m c_{m-1}^{*} \}
\\
f(s, m) 
= [(s+m)(s-m+1)]^{1/2}. 
\end{array}
\right.
\label{eqn:defA}
\end{equation}
By $\{ c_m \}$ we mean a set of the coefficients 
of the fiducial vector. 
We can easily verify (\ref{eqn:mat-pol}) 
by (\ref{eqn:CS1}) and (\ref{eqn:RS}). 
\par
The generating function for the general matrix elements exists 
as in the original CS \cite{Arec}. 
In the normal product form it reads 
\begin{eqnarray}
X_N (z_+, z_3, z_-) \equiv 
\left\langle {\bf\Omega}_2 \right\vert \, 
\exp(z_+ {\hat S_+}) \, \exp(z_3 {\hat S}_3) \, 
\exp(z_- {\hat S_-}) \, 
\left\vert {{\bf\Omega}_1} \right\rangle 
\nonumber \\
= \left\langle {\Psi_0} \right\vert {\hat R}^{+} ({\bf\Omega}_2) 
{\hat R}({\bf\Omega}) {\hat R}({\bf\Omega}_1) 
\left\vert {\Psi_0} \right\rangle 
= \left\langle {\Psi_0} \right\vert
 {\hat R}({\bf\Omega}^{''}) \left\vert {\Psi_0} \right\rangle. 
\label{eqn:genFunc}
\end {eqnarray}
where ${\bf\Omega}$ is related to $z_l\ (l = +, 3, -)$ 
through (\ref{eqn:EulerComp}); 
And ${\bf\Omega}^{''}$ is determined by (\ref{eqn:trirots}). 
Any matrix elements can be obtained from (\ref{eqn:genFunc})
via partial differentiations. 
\section{Path integral via the spin CS}
\label{sec:PI}
\subsection{Path integrals}
\label{sec:PI1}
In this section we will give the explicit path integral expression of the 
transition amplitude by means of the CS discussed in $\S$ \ref{sec:CS}. 
What we need is the propagator 
$K({\bf\Omega}_{f}, t_{f}; {\bf\Omega}_{i}, t_{i})$
 which starts from  $\left\vert {{\bf\Omega}_{i}} \right\rangle$ 
 at $t=t_{i}$, evolves 
 under the effect of the Hamiltonian 
${\hat H} (\hat S_{+}, \hat S_{-}, \hat S_{3}; t)$ 
which is assumed to be a function of $\hat S_{+}$, 
$\hat S_{-}$ and $\hat S_{3}$ with a suitable operator ordering 
and ends up with $\left\vert {{\bf\Omega}_{f}} \right\rangle$ 
at $t=t_{f}$: 
\begin{equation}
\fl 
K({\bf\Omega}_f, t_f; {\bf\Omega}_i, t_i)
=\left\langle {{\bf\Omega}_f, t_f} 
\vert 
{{\bf\Omega}_i, t_i} \right\rangle
=\left\langle {{\bf\Omega}_f} \right\vert \, 
T \exp [- (\i / \hbar) \int_{t_{j-1}}^{t_j} {\hat H}(t) \d t ] \, 
\left\vert {{\bf\Omega}_i} \right\rangle
\label{eqn:propagator1}
\end{equation}
where $T$ denotes the time-ordered product. 
The overcompleteness relation (\ref{eqn:polresolution}) affords us 
the well-known prescription of formal CSPI
 \cite{KlaSk,IKG} to give 
\begin{equation} 
K({\bf\Omega}_f, t_{f}; {\bf\Omega}_i, t_{i})
=\int  
\exp \{ (\i / \hbar) S[{\bf\Omega}(t)] \} \, 
{\cal D} [{\bf\Omega}(t)]
\label{eqn:PI}
\end{equation}
where
\begin{equation}
S[{\bf\Omega}(t)]
\equiv \int_{t_i}^{t_f} \d t \  
\Bigl[ \ \left\langle {\bf\Omega} \right\vert \i \hbar 
{\partial \over \partial t} 
 \left\vert {\bf\Omega} \right\rangle 
 - H({\bf\Omega}, t) \ \Bigr] 
 \equiv \int_{t_i}^{t_f}   
 L({\bf\Omega}, {\bf \dot \Omega},t) \d t 
\label{eqn:action-pol}
\end{equation}
and we symbolized
\begin{equation} 
{\cal D}[{\bf\Omega}(t)] 
\equiv
\lim_{N \rightarrow \infty} 
\prod_{j=1}^{N} \d{\bf\Omega}(t_j) 
=\prod_{t}\,  
[\sin\theta(t)\, \d \theta (t) 
\d \phi(t) \d \psi(t)]. 
\label{eqn:paths-pol}
\end{equation} 
The explicit form of the Lagrangian yields
\begin{equation}
\fl
L({\bf\Omega}, {\dot {\bf\Omega}}, t) 
= \hbar \Bigl[ A_0(\{ c_m \}) ({\dot \phi}\cos\theta + {\dot \psi})
 + A_3({\bf\Omega}, {\dot {\bf\Omega}}; \{ c_m \}) \Bigr] 
 - H({\bf\Omega}, t)
\label{eqn:Lagpol}
\end{equation}
where
\begin{eqnarray}
\fl
A_3({\bf\Omega}, {\dot {\bf\Omega}}; \{ c_m \}) 
\equiv 
-\frac12 
\sum_{m = -s+1}^{s} 
f(s, m) 
\Bigl[ ({\dot \phi} \sin\theta 
+ \i {\dot \theta}) \exp(\i \psi) c_m^{*} c_{m-1}
\nonumber \\ 
\lo \qquad \qquad \qquad
+ ({\dot \phi} \sin\theta - \i {\dot \theta}) 
\exp(- \i \psi)c_m c_{m-1}^{*} 
\Bigr].
\label{eqn:deffA3}
\end{eqnarray}
The term with square brackets in 
(\ref{eqn:Lagpol}) 
stemming from $\left\langle {\bf\Omega} \right\vert 
(\partial / \partial t) \left\vert {\bf\Omega} \right\rangle$ 
may be called the `topological term'
 that is related to the geometric phases 
in $\S$ \ref{sec:CSGP}.
\footnote{The significance of the term was once recognized by 
 Kuratsuji, who called it the `canonical term', 
 in relation to the semiclassical quantization; 
 note that the geometric phase associated with the term was called the 
 `canonical phase' in \cite{KMa} and \cite{KraC}; See \cite{KraC} 
 and references therein. We call them just the geometric phases 
 in the present paper.}
The proof of (\ref{eqn:Lagpol}) is best carried out 
by the use of the identity:
\begin{eqnarray}
\fl 
{\hat R}^{+}({\bf\Omega}) 
\frac{\partial}{\partial t} 
{\hat R}({\bf\Omega}) 
= {\hat R}^{+}({\bf\Omega})  
\Bigl( \, {\dot \phi} \frac{\partial}{\partial \phi}
+ {\dot \theta} \frac{\partial}{\partial \theta} 
+ {\dot \psi} \frac{\partial}{\partial \psi} 
\, \Bigr) {\hat R}(\bf\Omega) 
\nonumber \\
\fl=\i ({\dot \phi} \cos\theta + {\dot \psi}){\hat S_3}
+ \frac12 ( \i {\dot \phi} \sin\theta - {\dot \theta} ) 
\exp(\i \psi){\hat S_+} 
+ \frac12 ( \i {\dot \phi} \sin\theta + {\dot \theta} ) 
\exp(- \i \psi){\hat S_-}.
\label{eqn:pol-R}
\end{eqnarray}
and (\ref{eqn:mat-pol}). 
Since the relation (\ref{eqn:pol-R}) 
is independent of $s$, it can be readily verified 
by the use of a $2 \times 2$ matrix (\ref{eqn:s1/2}). 
\par
We have thus arrived at {\em the generic expressions 
of the path integrals via the SU(2)CS}, 
i.e. (\ref{eqn:PI})-(\ref{eqn:deffA3}), 
which constitute the main results of the present paper. 
We see from (\ref{eqn:Lagpol})-(\ref{eqn:deffA3}) 
that $\psi$ variable does not take effect when no neighbouring $\{ c_m \}$ 
exists for any $c_m$; 
Then $A_3$-terms vanishes and we can choose any $\psi$. 
%In this case CSPI is neatly described 
%in a pair of complex variable 
%we recover the result of \cite{spinCSPI} 
The special case when $c_m=1$ (for a sole $m$), 
which includes the conventional SU(2)CSPI, 
was once treated in \cite{PIspin}. 
The transition amplitude between any two states 
$\left\vert {i} \right\rangle$ 
at $t = t_{i}$ 
and $\left\vert {f} \right\rangle$ at $t = t_{f}$ can be evaluated by 
\begin{equation}
\int \!\!\! \int \d {\bf\Omega}_f 
\d {\bf\Omega}_i \ 
\left\langle {f} 
\vert 
{{\bf\Omega}_{f}} 
\right\rangle
K({\bf\Omega}_{f}, t_{f}; {\bf\Omega}_{i}, t_{i})
\left\langle {{\bf\Omega}_{i}} 
\vert {i} \right\rangle.
\label{eqn:tranamp}
\end{equation}
\subsection{Miscellaneous points}
In this subsection we study miscellaneous points concerning 
CSPI.
\par
First, we will investigate what information 
the semi-classical limit of CSPI brings. 
In the situation where 
$\hbar \ll S[{\bf\Omega}(t)]$, the principal contribution in (\ref{eqn:PI}) 
comes from the path that satisfies $\delta S =0$, which requires 
the Euler-Lagrange equations for 
$L({\bf\Omega}, {\bf \dot \Omega},t)$. 
Then we obtain
\begin{equation}
\fl 
\left
\{
\begin{array}{l}
\hbar 
\{ 
[A_0(\{ c_m \}) \sin\theta 
+ A_1(\psi; \{ c_m \}) \cos\theta 
] {\dot \phi} 
+ A_1(\psi; \{ c_m \}) {\dot \psi}
\} 
= - ( \partial H / \partial \theta )
\\
\hbar 
\{
[
A_0(\{ c_m \}) \sin\theta 
+ A_1(\psi; \{ c_m \}) \cos\theta 
] {\dot \theta} 
- [A_4(\psi; \{ c_m \}) \sin\theta] {\dot \psi}  
\} 
= \partial H / \partial \phi
\\
\hbar 
\{
[A_4(\psi; \{ c_m \}) \sin\theta] {\dot \phi}  
+ A_1(\psi; \{ c_m \}) {\dot \theta} 
\} 
= {\partial H / \partial \psi}
\label{eqn:vareq-pol} 
\end{array}
\right.
\end{equation}
where $A_0$ and $A_1$ are given by (\ref{eqn:mat-pol}) and 
\begin{equation}
\fl
A_4(\psi; \{ c_m \}) 
\equiv 
\frac{1}{2 \i} 
\sum_{m=-s+1}^{s} f(s, m) 
[ \exp(\i \psi) c_m^{*} c_{m-1} 
- \exp(- \i \psi) c_m c_{m-1}^{*} ]. 
\label{eqn:def-A4}
\end{equation}
Equations (\ref{eqn:vareq-pol}) 
are the generalized canonical equations. 
The special case, 
i.e. those for SU(2)CS with a fiducial vector 
$\left\vert {\Psi_0} \right\rangle = \left\vert {m} \right\rangle$, 
was treated in 
\cite{PIspin}; 
Putting $\left\vert {\Psi_0} \right\rangle 
= \left\vert {-s} \right\rangle$ 
brings us back to the results for the original case \cite{Klau,KUS}. 
\par 
Second, we point out that in the high spin limit 
the spin CS tends to the canonical CS $\left\vert {\alpha, n} \right\rangle$ 
\cite{NECSPI} 
as in the usual spin CS 
\cite{Rad,Arec,Perb}. 
And the results in $\S$ \ref{sec:CS} - $\S$ \ref{sec:CSGP} 
are easily converted 
to those of defined in 
\cite{NECSPI}; 
Especially, 
$A_3$-term approaches $A$-term 
of \cite{NECSPI}.
%Note that 
%it is necessary that $s$ and $m$ 
%are in the same order and have opposite signs 
%so that matrix elements may behave we%ll. 
%The condition is understood also as the requirement 
%for the finiteness of $n$.
\section{Applications to geometric phases} 
\label{sec:CSGP}
In this section we try to apply CS 
to the problems of geometric phases \cite{SW} 
to see what new effects the CS brings us. 
\subsection{Geometric phases}
\label{sec:GP}
Consider a cyclic change of CS, whose initial 
and final states are expressed as 
$\left\vert {{\bf\Omega}(0)} \right\rangle 
\equiv 
\left\vert {\phi(0), \theta(0), \psi(0)} \right\rangle$ 
and 
$\left\vert {{\bf\Omega}(t)} \right\rangle 
\equiv 
\left\vert {\phi(T), \theta(T), \psi(T)} \right\rangle$
 respectively, 
under the effect of a Hamiltonian $\hat H$ 
during the time interval $[0, T]$. 
In the light of the formalism developed in $\S$ \ref{sec:PI1}, 
the state vector accumulates the phase $\Phi(C)$ which amounts to
\footnote{It seems that there have been considerable works 
on CS and geometric phases. We do not intend to disregard them. 
  However, we site here \cite{KraC} 
  because it has cleared up the relation between CSPI 
and geometric phases and also because it is suitable 
for our later arguments. 
For an excellent survey and reprints 
on geometric phases see \cite{SW}.} 
\begin{equation}
 \Phi(C)
 = \left\langle {{\bf\Omega}(0)} 
 \vert 
 {{\bf\Omega}(t)} \right\rangle 
 = \exp \Bigl\{ (\i / \hbar) 
 \sum_{{\ssty {\rm all\ the\ cyclic}}
 \atop{\ssty {\rm paths}\ C}} 
\Bigl[ \Gamma(C) - \Delta(C) \Bigr]
\ \Bigr\}
\label{eqn:totph} 
\end{equation}
where
\begin{eqnarray}
\fl
\Gamma(C) 
\equiv \int_C \omega 
= \int_{0}^{T}  
\left\langle {\bf\Omega} \right\vert \, 
\i \hbar \, {\partial \over \partial t} 
\left\vert {\bf\Omega} \right\rangle \d t 
\nonumber \\
\lo= \int_{0}^{T}  
\hbar [A_0(\{ c_m \}) ({\dot \phi}\cos\theta + {\dot \psi})
+ A_3({\bf\Omega}, {\dot {\bf\Omega}}; \{ c_m \}) ] \d t
\label{eqn:Gam1}
\end{eqnarray}
is the `geometric phase' 
and 
\begin{equation}
\Delta(C) 
= \int_{0}^{T} \, \left\langle {\bf\Omega} \right\vert
{\hat H} \left\vert {\bf\Omega} \right\rangle \d t 
= \int_{0}^{T} \, H({\bf\Omega}, t) \d t.
\label{eqn:Del1}
\end{equation}
Note that, unlike the $A$-term in 
the canonical CSPI \cite{NECSPI}, 
$A_3$-term is not represented as a total derivative, 
and hence its effect does not vanish for a cyclic motion; 
It is only in the high spin limit that the effect disappears. 
\par 
We may describe the topological term in the 2-form: 
\begin{eqnarray}
\fl
\int_S  \d \omega
= \int_S 
\Bigl\{ - \hbar 
\Bigl[ 
\Bigl(
A_0(\{ c_m \}) \sin\theta 
+ A_1(\psi; \{ c_m \}) \cos\theta 
\Bigr) 
\d \theta \wedge \d \phi
\nonumber \\ 
\lo 
+ A_4(\psi; \{ c_m \}) \sin\theta 
\d \phi \wedge \d \psi 
+ A_1(\psi; \{ c_m \}) 
\d \psi \wedge \d \theta 
\Bigl]
\Bigr\}.
\label{eqn:2form-pol}
\end{eqnarray}
One may see that the strength of the well-known 
monopole-type term depends on $A_0$, i.e. 
the expectation value of the 
quantum number $m$ in the state of $\left\vert {\Psi_0} \right\rangle$. 
In addition we have {\em another field} 
with $A_1$ and $A_4$-terms describing the effect of 
interweaving coefficients of $\left\vert {\Psi_0} \right\rangle$ 
with their next ones.
\par 
In \cite{KraC} the path is chosen according 
to the variation principle 
in relation to the semiclassical 
approximation that plays a vital role 
in evaluating path integrals.  
And some developments of the CS geometric phases 
including the application in that direction has been done 
\cite{KMa,KMaa,KMb}. 
Yet such restriction is not necessary. 
Especially when the Hamiltonain is at most linear 
in the generators of the SU(2) algebra, however, 
the CS evolves as CS under the effect of the Hamiltonian 
and the resulting geometric phase agrees with that 
obtained by the variational path 
\cite{Perb,Perc}. 
 A somewhat deeper comment on the point can be found in \cite{NECSPI}. 
 We take the case for example in the following 
discussion, partly because we want to compare the result with 
that obtained before 
\cite{KMa} 
and investigate the effect of CS, 
and partly because the method of CSPI gives us 
a clear insight even into the case; 
We will see that the expressions 
(\ref{eqn:PI})-(\ref{eqn:deffA3}) 
and (\ref{eqn:2form-pol}) 
indeed give us a key to find the effects 
in $\S$ \ref{sec:CSmodel}. 
\subsection{The model systems} 
\label{sec:CSmodel} 
We will apply the formalism in $\S$ \ref{sec:GP} 
to model systems. 
As stated earlier we can choose any $\psi$ 
in $\S$ \ref{sec:simplest} - $\S$ \ref{sec:simp-s1}; 
We put $\psi = - \phi$ right from the start in these subsections 
so as to proceed parallel to \cite{KMa}. 
\subsubsection{The simplest case}
\label{sec:simplest}
First, we take up the system in \cite{KMa}. 
Consider a particle with spin $s$ in a time-dependent 
magnetic field given by 
\begin{equation}
{\bf B}(t)=(B_0 \cos \omega t,B_0 \sin \omega t,B). 
\label{eqn:Bfield}
\end{equation}
${\bf B}(t)$ consisits of a static field along the $z$-axis 
plus a time-dependent 
rotating one perpendicular to it with a frequency $\omega$, 
which is familiar in magnetic resonance. 
The Hamiltonian is given by 
\begin{equation}
\hat H(t) =-\mu {\bf B}(t) \, {\bf \cdot} 
\, {\bf {\hat S}}.
\label{eqn:Bhaml1}
\end{equation}
We assume that the system be described by the spin CS 
$\left\vert {\bf\Omega} \right\rangle$ 
with $\left\vert {\Psi_0} \right\rangle = \left\vert {m} \right\rangle$. 
Thus there is no $A_3$-term in the Lagrangian. 
Since the Hamiltonian is written in the linear combination 
of the generators of SU(2) algebra, we only need to consider 
the variation path in the present case. 
\par 
Now, following $\S$ \ref{sec:GP}, the expectation value 
of the Hamiltonian is 
\begin{equation}
H(\theta, \phi) 
= - \mu m \, [ \, B_0 \sin \theta \cos (\phi-\omega t) 
+ B \cos \theta \, ]. 
\label{eqn:Bhaml2}
\end{equation}
Care should be taken that the phase convention is different 
from that in 
\cite{KMa}: in the latter the sign of $\theta$ is reversed. 
Then from (\ref{eqn:vareq-pol}), 
we have the variation equation 
\begin{equation}
\dot\theta 
= (\mu B_0 / \hbar) \, \sin (\phi - \omega t ) 
\qquad
\dot \phi 
= (\mu / \hbar) \, 
[ \, B_0 \cot \theta \cos ( \phi - \omega t ) - B \, ]. 
\label{eqn:Bvareq}
\end{equation}
One sees that this form of equations of motion allows 
a special cyclic solution with the period 
$T = 2 \pi / \omega$ \cite{KMa}:
\begin{equation}
\phi = \omega t \qquad \theta = \theta_0 (= {\rm const} )
\label{eqn:reseq}
\end{equation}
where the following relation should hold among 
a set of parameters $\theta_0$ and $(B, B_0, \omega)$:
\begin{equation}
\cot \theta_0 
= (B /B_0) + [\hbar \omega / (\mu B_0)].
\label{eqn:rescond}
\end{equation}
Equation (\ref{eqn:reseq}) represents a cyclic path $C$ 
on the $(\theta, \phi)$-space, i.e., a unit sphere. 
\par 
Next we turn to the evaluation of the geometric phase $\Gamma(C)$ 
for (\ref{eqn:reseq}). 
From (\ref{eqn:Gam1}) it becomes 
\begin{equation}
\Gamma(C)= - 2\pi m \hbar  (1- \cos \theta_0) 
=  - m \hbar \Omega(C) 
\label{eqn:Gam2}
\end{equation}
$\Omega(C)$ being the solid angle subtended 
by the curve $C$ at the origin of the spin phase space. 
On the other hand, from (\ref{eqn:Del1}) 
the Hamiltonian phase $\Delta(C)$ is given by 
\begin{equation}
\Delta(C)= - \frac{2 \pi \mu m}{\omega} 
\, ( B \cos \theta_0 + B_0 \sin \theta_0 ). 
\label{eqn:Del2}
\end{equation}
\par 
In order to detect the phase $\Gamma(C)$ experimentally 
we invoke the interference phenomena of two beams 
\cite{KMa}. 
Imagine a particle beam that is made up of atoms with spin-$s$ 
and in the state of $\left\vert {\theta, \phi} \right\rangle$. 
Assume that the beam is split into two parts, 
one of which is subjected to a magnetic field in 
(\ref{eqn:Bfield}) and the other is not. 
And besides, we assume that the controlling parameters 
$(B_0, B, \omega)$ satisfy the condition: $\Delta(C) = 0$. 
Then these two beams are 
designed to come into reunion after a time interval $T$. 
Under such situation (\ref{eqn:totph}) shows 
that the resultant interference pattern 
is determined solely by the geometric phase; 
its maximum intensity $I$ goes as 
\begin{equation}
I \propto  1 + \cos [{\Gamma(C)}/{\hbar}]. 
\label{eqn:intfer}
\end{equation}
Let us see the situation more specifically. 
The vanishing of $H(t)$ is sufficient for that of $\Delta(C)$. 
From (\ref{eqn:Bhaml2}) it occurs when  
$\cot \theta_0 = - B_0 / B$. 
By combining this with (\ref{eqn:rescond}), 
we get 
$
\omega = -\mu \, (B_0^2 +B^2) / (\hbar B).
$ 
So the interference pattern depends upon 
\begin{equation}
   \Gamma(C) = - 2m \pi [1 + B_0 {(B_0^2 + B^2)}^{-1/2}].
\label{eqn:Gam3}
\end{equation}
Specifically when $m=0$, we find $\Gamma(C)$ disappears. 
\par
With the conventional choice of a fiducial vector, 
SU(2)CSPI merely gives the result for 
$c_{-s}=1$ with other members of $\{ c_m \}$ vanished, 
i.e. $m = -s$  
in (\ref{eqn:Gam2})-(\ref{eqn:Gam3}) 
\cite{KMa}. 
Contrary to this, 
we have here obtained a stronger result (\ref{eqn:Gam2}) which, 
in an adiabatic case, reduces to that of \cite{Berry}. 
Thus the formula (\ref{eqn:2form-pol}) 
and the argument just after them reveal that 
{\em the SU(2)CSPI is 
mature and complete enough to incorporate the formula 
{\rm (\ref{eqn:Gam2})} 
as its special case 
and moreover it also covers the wider cases as described 
in the following subsections}. 
From our viewpoint it is clear why we encounter 
the monopole formula so often in geometric phases: 
it is a natural consequence of the fact that the physically 
important systems are sometimes described by the SU(2) CS. 
\subsubsection{A simple case with spin 1}
\label{sec:simp-s1} 
Consider a spin-1 particle prepared in the CS 
$\left\vert {\phi_0, \theta_0, \psi_0 = -\phi_0} \right\rangle$ 
with 
$
\left\vert {\Psi_0} \right\rangle = (\frac23)^{1/2} 
\left\vert {1} \right\rangle 
+ (\frac13)^{1/2}\left\vert {-1} \right\rangle
$
and the same magnetic field as (\ref{eqn:Bfield}). 
Notice that this fiducial vector 
cannot be obtained from $\left\vert {m} \right\rangle\ (m = -1, 0, 1)$ 
by $R^{(1)}({\bf\Omega})$. 
Since $A_0 = \frac13$ and $A_3$-term vanishes, 
all the equations in $\S$ \ref{sec:simplest} for $m = 1$ hold 
if we replace $m$ with $\frac13$; 
And the result gives 
\begin{equation}
\Gamma(C) = \frac23 \pi (\cos\theta_0 - 1)
\label{eqn:GPspin1}
\end{equation}
which clearly differs from the $m=1$ or $m=-1$ case in 
(\ref{eqn:Gam2}); 
we can distinguish one from the other by the experiment 
discussed in $\S$ \ref{sec:simplest}.
\subsubsection{A special case with spin 1}
\label{sec:spec-s1}
We next treat a spin-1 particle prepared in the spin CS 
$\left\vert {\phi_0, \theta_0, \psi_0} \right\rangle$ with 
$
\left\vert {\Psi_0} \right\rangle = (\frac12)^{1/2} 
\left\vert {1} \right\rangle 
+ (\frac12)^{1/2}\left\vert {-1} \right\rangle.
$ 
Then, since 
$A_0=A_3=0$ and $A_1 = A_2 = 0$, 
we always have
\begin{equation}
\Gamma(C) = \Delta(C) = 0
\end{equation}
for arbitrary magnetic fields. 
Consequently, we find no change of interference fringes 
produced by the two split beams 
like those described in $\S$ \ref{sec:simplest} 
even though one of which is subjected 
to any varying magnetic fields. 
\par
It is readily seen that we are brought 
into the same situation 
when we deal with a spin-$s$ particle 
which is prepared in the CS with $\left\vert {\Psi_0} \right\rangle$ 
fulfilling the condition that 
no neighbouring $\{ c_m \}$ exists for any $c_m$ 
as well as $\vert {c_m} \vert = \vert {c_{-m}} \vert$. 
\subsubsection{Manifest appearance of the $A_3$-term}
\label{sec:manif-s1}
In the last case we deal with a spin-$1$ particle 
in the CS 
$\left\vert {\phi_0 = 0, \theta_0 = {\rm const}, \psi_0 = 0} \right\rangle$
 having a fiducial vector 
$
\left\vert {\Psi_0} \right\rangle = (\frac13)^{1/2} 
\left\vert {1} \right\rangle 
+ (\frac13)^{1/2} \left\vert {0} \right\rangle 
+ (\frac13)^{1/2} \left\vert {- 1} \right\rangle
$
that cannot be obtained from 
$\left\vert {m} \right\rangle\ (m = -1, 0, 1)$ by 
$R^{(1)}({\bf\Omega})$. 
Then, since $A_0$-term vanishes, 
we can see what $A_3$-term brings us explicitly. 
\par
Now, let us revisit the system in $\S$ \ref{sec:simplest}; 
We have the Hamiltonian (\ref{eqn:Bhaml1}) with a magnetic field 
(\ref{eqn:Bfield}). 
Then we have 
\begin{equation}
\fl 
\begin{array}{l}
H({\bf\Omega}) 
= - \frac{1}{3}\sqrt{2} \mu 
\{ B_0 [ (1 + \cos\theta) \cos(\phi - \omega t + \psi) 
- (1 - \cos\theta) \cos(\phi - \omega t - \psi) ] 
\\
 \qquad \qquad - 2 B \cos\psi \sin\theta \}
\label{eqn:Bhaml3}
\end{array}
\end{equation}
and it follows readily from the variation equations 
(\ref{eqn:vareq-pol}) that 
\begin{equation}
\fl 
\left
\{
\begin{array}{l}
\{ 
(\dot\phi \cos\theta + \dot\psi)
+ (\mu / \hbar) 
[B_0 \sin\theta \cos(\phi - \omega t) + B \cos\theta]
\} 
\cos\psi = 0
\\
{\dot \theta} \cos\psi \cos\theta  
- {\dot \psi} \sin\psi \sin\theta 
\\
\qquad = [\mu B_0 / (2 \hbar)] 
[ (1 + \cos\theta) \sin(\phi - \omega t + \psi) 
    - (1 - \cos\theta) \sin(\phi - \omega t - \psi) ]
\\
{\dot \phi} \sin\psi \sin\theta 
+ {\dot \theta} \cos\psi 
\\
\qquad =  [\mu / (2 \hbar)] 
\{  B_0 [ (1 + \cos\theta) \sin(\phi - \omega t + \psi) 
    + (1 - \cos\theta) \sin(\phi - \omega t - \psi) ] 
\\
\lo \qquad \qquad \qquad - 2 B \sin\psi \sin\theta \}. 
\label{eqn:Bvareq2} 
\end{array}
\right.
\end{equation}
It follows that (\ref{eqn:Bvareq2}) allows 
a special cyclic solution with the period 
$T = 2 \pi / \omega$:
\begin{equation}
\phi = \omega t 
\qquad
\theta = \theta_0
\qquad
\psi = 0
\label{eqn:Bvar-sol}
\end{equation}
where the relation
\begin{equation}
\tan\theta_0 = - \{ (B /B_0) + [\hbar \omega / (\mu B_0)] \}
\label{eqn:rescond2}
\end{equation} 
should hold among a set of parameters $\theta_0$ and $(B, B_0, \omega)$. 
The solution (\ref{eqn:Bvar-sol}) describes a cyclic trajectory
 $C$ in the $\bf\Omega$-space, 
whose projection on the $(\theta, \phi)$-space, i.e. a unit sphere, 
is a cone with the origin as its vertex. 
Then we have from (\ref{eqn:Gam1}) 
\begin{equation}
\Gamma(C) = - \frac{4 \sqrt{2}}{3} \hbar \pi \sin\theta_0
\label{eqn:GPmanif-app1}
\end{equation} 
and from (\ref{eqn:mat-pol}) and (\ref{eqn:Del1})
\begin{equation}
\Delta(C) = - \frac{4 \sqrt{2}}{3} 
\frac{\pi \mu}{\omega} 
( B_0 \cos\theta_0 - B \sin\theta_0 ). 
\label{eqn:Delmanif-app}
\end{equation}
We can derive $\Gamma(C)$ also from 2-form (\ref{eqn:2form-pol}). 
\par 
To see effect of the geometric phase solely, 
we consider the same experiment as $\S$ \ref{sec:simplest}; 
we prepare two split beam in one of which 
$H(t) = 0$. The condition is sufficient for $\Delta(C) =0$. 
From (\ref{eqn:Bhaml3}) this happens when 
$\tan\theta_0 = B_0/B$, which, with the aid of (\ref{eqn:rescond2}), 
gives 
$
\omega = -\mu \, (B_0^2 +B^2) / (\hbar B)
$. 
Next we recombine the beam with another split beam 
with no fields experienced. 
Then the interference fringes depends upon
\begin{equation}
\Gamma(C) = - \frac{4 \sqrt{2}}{3} \hbar \pi [ B_0 (B_0^2 + B^2)^{- 1/2} ].
\label{eqn:GPmanif-app2}
\end{equation}
\par
In general we have a similar result 
for a spin-$s$ particle in the CS with $\left\vert {\Psi_0} \right\rangle$ 
when $\vert {c_m} \vert = \vert {c_{-m}} \vert$ holds for any $m$. 
The spin-$\frac12$ case is rather trivial since any fiducial vector 
$\left\vert {\Psi_0} \right\rangle$ can be reached from 
$\left\vert {- \frac12} \right\rangle$ 
by ${\hat R}^{(1/2)}({\bf\Omega})$, 
thus yielding a re-parametrization of ${\bf\Omega}$ 
in the conventional CS. 
\section{Summary and prospects}
\label{sec:summary} 
We have investigated a natural extension of the spin or SU(2)CSPI, 
which turns out to be performed successfully. 
\par
Conventional canonical CS and spin CS 
have been playing the roles of macroscopic wave functions 
in vast fields from lasers, superradiance, superfluidity 
and superconductivitiy to nuclear and particle physics 
\cite{KlaSk}. 
Thus we may expect that by choosing approapriate sets of $\{ c_m \}$ 
the CS evolving from arbitrary fiducial vectors 
will serve as approximate states 
or trial wave functions for the collective motions, 
having higher energies in 
various macroscopic or mesoscopic quantum phenomena 
such as spin vortices \cite{KraYab} and domain walls, 
which may not be treated by the former. 
We hope that numerous applications of the CS and CSPI 
will be found in the near future. 
\par
From the viewpoint of mathematical physics, 
it is desirable that the present CSPI formalism 
is extended to more wider classes. 
The generalization to the SU(1, 1) CS case, 
which is closely related to squeezed states 
in lightwave communications and quantum detections 
\cite{SQ}, 
is one of the highly probable candidates 
and may be treated elsewhere. 
\ack 
The author sincerely thanks Prof T Sakuragawa 
and the colleagues of the mathematical information group 
at Kyoto University concerning the electronic submission. 
He also expresses his deep appreciation to 
the members of Department of Physics 
at Ritsumeikan University-BKC for their warm hospitality 
where part of the present work has been performed; 
special thanks are to Prof H Kuratsuji for his comments on 
the earlier version of the manuscript \cite{spinCSPI}
 and encouragement. 
\appendix
\section{Some formulae for rotation matrices}
\label{sec:Rots}
Some basic formulae on the properties of the rotation matrices 
are enumerated \cite{Mess,BS,Miller}. 
 We are in need of them in $\S$ \ref{sec:CS}. 
We mainly follow the notation of Messiah \cite{Mess}. 
\\
 (i) {\em Matrix elements}\\
A rotation with Euler angle $(\phi, \theta, \psi)$ 
of a spin-$s$ particle is 
specified by a matrix $\hat R$:
$
{\hat R}(\phi, \theta, \psi) 
= \exp(-\i\phi {\hat S_3}) \exp(-\i\theta {\hat S_2}) 
\exp(-\i\psi {\hat S_3})
$ 
which is a $(2s+1) \times (2s+1)$ matrix whose components 
$ 
R_{m{}m'}^{(s)}(\phi, \theta, \psi)
\equiv \left\langle {s, m} \right\vert {\hat R}(\phi, \theta, \psi) 
\left\vert {s, m'} \right\rangle
$
are 
$
R_{m{}m'}^{(s)}(\phi, \theta, \psi) 
= \exp(-\i\phi m) \, r_{m{}m'}^{(s)}(\theta) \, 
\exp(-\i\psi m').
$
Here 
$
r_{m{}m'}^{(s)}(\theta)
\equiv \left\langle {s, m} \right\vert \exp(-\i\theta {\hat S_2}) 
\left\vert {s, m'} \right\rangle
$ 
is determined by the Wigner formula \cite{Mess}. 
In particular, if $s=\frac12$, 
$R_{m{}m'}^{(s)}$ is extremely simple to give: 
\begin{equation}
\fl
{\hat R}^{(1/2)}(\phi, \theta, \psi) 
= \left(
\begin{array}{lc}
\cos( \case12 \theta ) 
\exp[- \case12 \i (\phi + \psi)] 
& -\sin( \case12 \theta ) 
\exp[- \case12 \i (\phi - \psi)]
\\
\sin( \case12 \theta ) \exp[ \case12 \i (\phi - \psi)] 
& \cos( \case12 \theta ) \exp[ \case12 \i (\phi + \psi)] 
\end{array}
\right).
\label{eqn:s1/2}
\end{equation}
Most of the following relations, being independent of $s$, 
can be readily verified by the use of (\ref{eqn:s1/2}). 
\\
(ii) {\em Gaussian decomposition} \cite{Arec,Perb,Perc,Gil}\\
The rotation matrix ${\hat R}({\bf\Omega})$ can be 
put into the normal or anti-normal ordering form 
in which ${\hat R}$ is 
specified by a set of complex variables:
\begin{eqnarray}
\fl 
{\hat R}(\phi, \theta, \psi) 
= {\hat R}(z_+, z_3, z_-)
\equiv \exp(z_+ {\hat S_+}) \exp(z_3 {\hat S_3}) 
\exp(z_- {\hat S_-})
\nonumber \\
\lo= \exp(z_- {\hat S_-}) \exp(- z_3 {\hat S_3}) 
\exp(z_+ {\hat S_+}) 
. 
\label{eqn:compRot}
\end{eqnarray}
The relation between the Euler angles and the complex parameters 
is given by
\begin{equation}
\left\{
\begin{array}{l}
z_+ = - \tan ( \case12 \theta ) \exp(- \i \phi) 
\nonumber \\
z_3 = - 2 \ln \{ \cos( \case12 \theta ) 
\exp[- \i (\phi + \psi)] \} \\
z_- = \tan ( \case12 \theta ) \exp(- \i \psi) .
\nonumber \\  
\end{array}
\right.
\label{eqn:EulerComp}
\end{equation} 
(iii) {\em Combinations with $\hat {\bf S}$} \cite{Mess}
\begin{equation}
\fl 
\left\{
\begin{array}{l}
{\hat R^{+}}({\bf\Omega}) {\hat S}_3 {\hat R}({\bf\Omega}) 
 = \cos\theta {\hat S}_3 
 - \frac12 \sin\theta 
 [ \exp(\i \psi) {\hat S}_{+} + \exp(-\i \psi) {\hat S}_{-} ]
 \\
{\hat R^{+}}({\bf\Omega}) {\hat S}_{\pm} {\hat R}({\bf\Omega})
= \sin\theta \exp(\i \phi) {\hat S}_3 
\\
\qquad \qquad 
+ \frac12 
\{
(\cos\theta \pm 1) \exp[\i (\pm \phi + \psi)]{\hat S}_{+} 
+ (\cos\theta \mp 1) \exp[\i (\pm \phi - \psi)]
{\hat S}_{-}
\}. 
\end{array}
\right.
\label{eqn:RS}
\end{equation}
(iv) {\em Inverse}\\
$\hat R$ is unitary and its inverse matrix is given by 
\begin{equation}
{\hat R}^{+}(\phi, \theta, \psi) 
= {\hat R}^{-1}(\phi, \theta, \psi)
= {\hat R}(-\psi, -\theta, -\phi). 
\label{eqn:rotinverse}
\end{equation} 
(v) {\em Orthogonality relation}\\
The relation stems from integrating the products of the unitary 
irreducible representaions of a compact groups 
over the element of the group; thus it is a generic relation 
for the representations. In the present case it reads 
\cite{BS,Miller}
\begin{equation}
\fl 
\int_0^{2\pi} \int_0^{\pi} \int_0^{2\pi}\ 
\left( R_{m{}m'}^{(s)}(\phi, \theta, \psi) \right)^{*} 
R_{n{}n'}^{(s')}(\phi, \theta, \psi) 
\sin\theta\ \d\phi \d\theta \d\psi 
= \frac{8\pi^{2}}{2s+1}\ 
\delta_{m, n} \delta_{m', n'} \delta_{s, s'}.
\label{eqn:ortho}
\end{equation}
(vi) {\em Two successive rotations}\\
Two succesive rotations specified by Euler angles 
${\bf\Omega}_l \equiv (\phi_l, \theta_l, \psi_l) \  (l = 1, 2)$ 
produce 
$
{\hat R}({\tilde {\bf\Omega}}) 
\equiv 
{\hat R}({\bf\Omega}_2) {\hat R}({\bf\Omega}_1) 
$ 
where ${\tilde {\bf\Omega}} \equiv ({\tilde \phi}, {\tilde \theta}, 
{\tilde \psi})$ obeys 
\begin{equation}
\fl 
\left\{
\begin{array}{l}
\cos{\tilde \theta} = \cos\theta_1 \cos\theta_2 
- \sin\theta_1 \sin\theta_2 \cos(\phi_1 + \psi_2) \\
\sin{\tilde \theta} \exp(\i {\tilde \phi}) \nonumber \\
= \exp(\i \phi_2) \, \Bigl[ \cos\theta_1 \sin\theta_2 
+ \sin\theta_1 \cos\theta_2 \cos(\phi_1 + \psi_2) 
+ \i \sin\theta_1 \sin(\phi_1 + \psi_2) \Bigr] \\
\cos( \case12 {\tilde \theta} ) 
\exp[ \case12 \i (\tilde \phi + \tilde \psi)] 
 \\
= \exp[ \case12 \i (\phi_2 + \psi_1) ] 
 \Bigl\{ 
\cos( \case12 \theta_1 ) \cos( \case12 \theta_2) 
\exp[ \case12 \i (\phi_1 + \psi_2) ] 
\\
\qquad - \sin( \case12 \theta_1 ) 
\sin( \case12 \theta_2 ) 
\exp[ -\case12 \i (\phi_1 + \psi_2) ] \Bigr\}.
\label{eqn:tworots}
\end{array}
\right.
\end{equation}
It is obvious that (\ref{eqn:tworots}) above 
and (\ref{eqn:trirots}) below are the relations 
from the spherical trigonometry. \\
(vii) {\em Three successive rotations}\\
In a similar manner to that in (vi), the Euler angles made of 
three successive rotations can be calculated. 
Assuming that the rotations are specified by 
Euler angles 
$(\phi_1, \theta_1, \psi_1)$, $(\phi, \theta, \psi)$ 
and $(\phi_2, \theta_2, \psi_2)$, which happen in this order, 
the composed rotation yields 
$
{\hat R}({\bf\Omega}') 
\equiv 
{\hat R}({\bf\Omega}_2)
{\hat R}({\bf\Omega})
{\hat R}({\bf\Omega}_1) 
$
where ${\bf\Omega}' \equiv (\phi', \theta', \psi')$ obeys  
\begin{eqnarray}
\fl 
\cos{\theta'} = [ \cos\theta_1 \cos\theta 
- \sin\theta_1 \sin\theta \cos(\phi_1 + \psi) 
]\cos\theta_2 
+ \{ \sin\theta_1 [ \sin(\phi_1 + \psi) 
\sin(\phi + \psi_2) 
 \nonumber \\ 
\lo
- \cos(\phi_1 + \psi) \cos\theta \cos(\phi + \psi_2) ] 
- \cos\theta_1 \sin\theta \cos(\phi + \psi_2) 
\} \sin\theta_2
\label{eqn:trirots}
\end{eqnarray}
and two additional equations that we omit here; 
They describe $\sin\theta' \exp(\i \phi')$ 
and $\cos(\frac12 \theta') 
\exp[\frac12 \i (\phi' + \psi')]$ in terms of 
${\bf\Omega}_1$, ${\bf\Omega}$ 
and ${\bf\Omega}_2$.
%%%
%
%@@@@@@@@@@@@@@@@@@@@@@@@@@@@@
%\newpage

%@@@@@@@@@@@@@@@@@@@@@@@@@@@ 
\end{document}